\documentclass[12pt,preprint]{aastex}

\begin{document}

\title{Origin of the galaxy morphology}
\author{N.Nakasato\altaffilmark{1}}
\altaffiltext{1}{Astronomisches Rechen-Institut, M\"ochhofstrasse 12-14, 69120 Heidelberg, Germany}

\section{Introduction}
In the current most plausible Cold Dark Matter (CDM) cosmology, larger halos increase their mass
by the progressive mergers of smaller clumps. 
Due to these progressive merger events, galaxies have formed and evolved.
Such merger events could trigger star bursts depending on mass of a merging object.
In other words, star formation history reflects the strength
of the interaction between a galaxy and merging objects.
Also, a several merger events strongly affect
the development of the morphology of galaxies as assumed in semi-analytic models.
In the most advanced semi-analytic models, N-body simulations of dark matter particles are
used to obtain the merging history of halos. 
By combining the description of radiative cooling, hydrodynamics and star formation
with the obtained merging history, such models successfully have explained the various qualitative predictions.
Here, we show the results of similar approach but using a fullly numerical model.
In contrast to the semi-analytic models, we use our high resolution
Smoothed Particle Hydrodynamics (SPH) models.
With our SPH code, we try to tackle the problem of the galaxy morphology.
We have done a several handful high-resolution SPH simulations
and analyzed the merging history of such models.
Accordingly, we can see the relation between the obtained morphology and the merging
history or other physical properties of the model.

\section{Method and Initial Models}
In the CDM cosmology, the early stage of the formation of galaxies involve the progressive merger
of sub-galactic clumps.
To properly model the dynamics of the merging history, we adopt the three-dimensional
SPH method.
Main features of our SPH code are (1) N-body integrator using GRAPE,
(2) metallicity dependent cooling function, (3) a star formation algorithm and (4) a self-consistent
treatment of chemical enrichment history of interstellar matter (ISM). 
For initial models, we set up 3 $\sigma$ over-density sphere
with the small scale CDM power spectrum.
We assume the standard CDM model, the mass of the sphere is $\sim$ $10^{12}$ $M_{\odot}$,
and the baryon fraction is 0.1.
We produce 36 such sphere using different seeds for the random number generator.
We evolve each model from redshift (z) $\sim$ 25 to $\sim$ 1.
Each model starts with the number of particles (N) $\sim$ 35,000 - 55,000
(a half for gas particles and others for dark matter particles) and ends
with N $\sim$ 70,000 - 110,000 (N$_{star}$ $\sim$ 42,000 - 76,000). 
These differences among models depend on the seed for the random number generator.

\section{Results}
By examining the projected density at z $\sim$ 1, 
we can classify each model with the its appearance.
Three major types are A) a spiral galaxy, B) an elliptical galaxy, and
C) multiple galaxies.
Each type shows different star formation history (merging history) as shown in Figure 1.
There are 8 galaxies which are classified as the type-A.
Evolution of star formation rates (SFR) for these model
have a strong peak in the early stage of evolution (the left panel).
The type-B models show two peaks in the SFR history (the center panel).
Moreover, the type-C models typically show more than two peaks (the right panel).
One such model has two close dwarf galaxies and shows second and third peaks in
the SFR plot. By examining the evolution of the particle orbit, these second and third
peaks are caused by close encounters with dwarfs. 
At this time, the long interval of the output of the particles data prevent
us to do a detailed analysis of the merging history.
We will fix this problem in the future work.

As an example of other application of our approach, 
we plot compute evolution of global history of our model in Figure 2.
In the left panel, the history of [Fe/H] as a function of redshift is plotted.
The solid ane dashed lines correspond to the metallicity for star and gas particles
and for only gas particles, respectively.
The global star formation history is presented in the right panel. 
Since all 36 models used here start from high density peak, 
the peak SFR occurs much higher than expected.
By increasing the number of models and classifying the model
with a more systematic way, we can plot the global star formation
for different type of galaxies in the future work.

\clearpage

\begin{figure}
\plotone{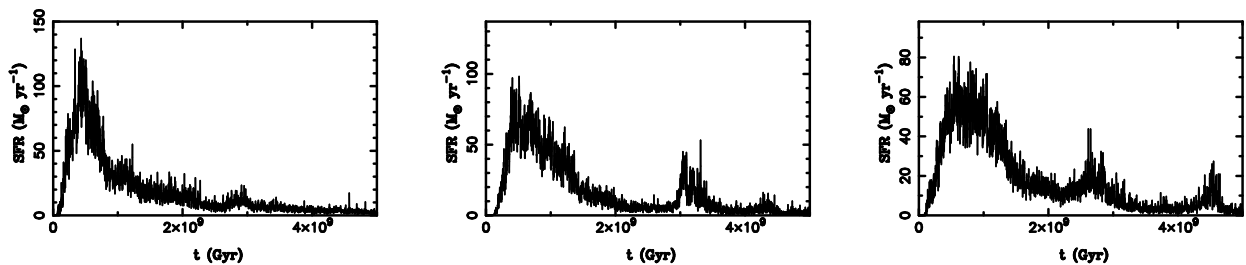}
\caption{Three examples of star formation history. From the left to the right, type-A, type-B and type-C.}
\end{figure}

\begin{figure}
\plotone{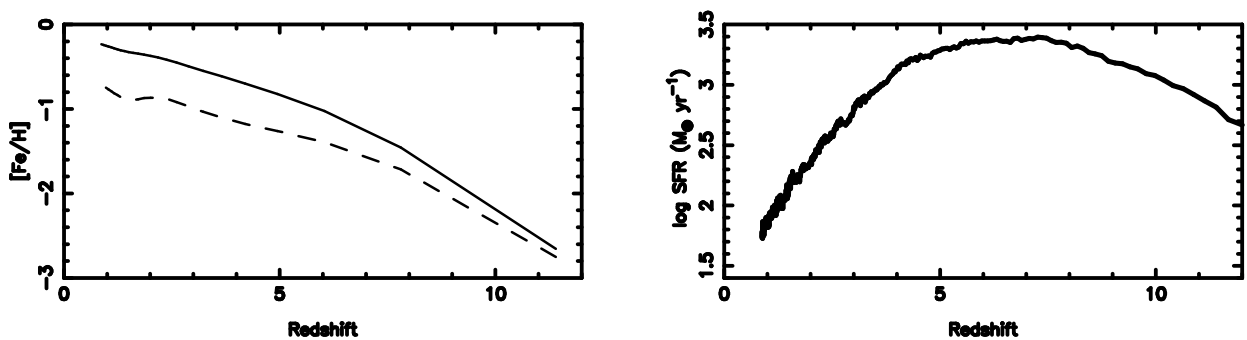}
\caption{The global history of [Fe/H] and SFR for all our model}
\end{figure}

\end{document}